\documentclass[12pt,preprint]{aastex}
\shorttitle{Weak Lensing By Coma}
\shortauthors{Kubo et al.}

\begin{document}

%% LaTeX will automatically break titles if they run longer than
%% one line. However, you may use \\ to force a line break if
%% you desire.

\title{The Mass Of The Coma Cluster From Weak Lensing In The Sloan Digital Sky Survey}

%% Use \author, \affil, and the \and command to format
%% author and affiliation information.
%% Note that \email has replaced the old \authoremail command
%% from AASTeX v4.0. You can use \email to mark an email address
%% anywhere in the paper, not just in the front matter.
%% As in the title, use \\ to force line breaks.

\author{Jeffrey M. Kubo\altaffilmark{1}, Albert Stebbins\altaffilmark{1}, James Annis\altaffilmark{1},  Ian P. Dell'Antonio\altaffilmark{2}, Huan Lin\altaffilmark{1}, Hossein Khiabanian\altaffilmark{2}, Joshua A. Frieman \altaffilmark{1,3}}
\altaffiltext{1}{Center for Particle Astrophysics, Fermi National Accelerator Laboratory, Batavia, IL 60510}
\altaffiltext{2}{Physics Department, Brown University, Box 1843, Providence, RI 02912}
\altaffiltext{3}{Kavli Institute for Cosmological Physics and Department of Astronomy \& Astrophysics, University of Chicago, Chicago, IL 60637}

%% Notice that each of these authors has alternate affiliations, which
%% are identified by the \altaffilmark after each name.  Specify alternate
%% affiliation information with \altaffiltext, with one command per each
%% affiliation.

%% Mark off your abstract in the ``abstract'' environment. In the manuscript
%% style, abstract will output a Received/Accepted line after the
%% title and affiliation information. No date will appear since the author
%% does not have this information. The dates will be filled in by the
%% editorial office after submission.

\begin{abstract}
We present a weak lensing analysis of the Coma Cluster using the Sloan Digital Sky Survey (SDSS) Data Release Five.  Complete imaging of a $\sim200$ square degree region is used to measure the tangential shear of this cluster.  The shear is fit to an NFW model and we find a virial radius of $r_{200}=1.99_{-0.22}^{+0.21}\mathrm{h^{-1}Mpc}$ which corresponds to a virial mass of $M_{200}=1.88_{-0.56}^{+0.65}\times10^{15}\mathrm{h^{-1}M_{\odot}}$.  We additionally compare our weak lensing measurement to the virial mass derived using dynamical techniques, and find they are in agreement.  This is the lowest redshift, largest angle weak lensing measurement of an individual cluster to date.
\end{abstract}

%% Keywords should appear after the \end{abstract} command. The uncommented
%% example has been keyed in ApJ style. See the instructions to authors
%% for the journal to which you are submitting your paper to determine
%% what keyword punctuation is appropriate.


%% From the front matter, we move on to the body of the paper.
%% In the first two sections, notice the use of the natbib \citep
%% and \citet commands to identify citations.  The citations are
%% tied to the reference list via symbolic KEYs. The KEY corresponds
%% to the KEY in the \bibitem in the reference list below. We have
%% chosen the first three characters of the first author's name plus
%% the last two numeral of the year of publication as our KEY for
%% each reference.

%% Authors who wish to have the most important objects in their paper
%% linked in the electronic edition to a data center may do so by tagging
%% their objects with \objectname{} or \object{}.  Each macro takes the
%% object name as its required argument. The optional, square-bracket
%% argument should be used in cases where the data center identification
%% differs from what is to be printed in the paper.  The text appearing
%% in curly braces is what will appear in print in the published paper.
%% If the object name is recognized by the data centers, it will be linked
%% in the electronic edition to the object data available at the data centers
%%
%% Note that for sources with brackets in their names, e.g. [WEG2004] 14h-090,
%% the brackets must be escaped with backslashes when used in the first
%% square-bracket argument, for instance, \object[\[WEG2004\] 14h-090]{90}).
%%  Otherwise, LaTeX will issue an error.

\section{Introduction}
Measurements of the mass of the Coma cluster date back to \citet{zwicky33} who first applied the virial theorem to Coma and showed that dark matter dominates the cluster on megaparsec scales. Since then a number of methods have been used to determine the mass of Coma under various assumptions.  \citet{kent82} used the galaxy surface density profile to measure the mass under the assumption that the mass traces the galaxy distribution.  \citet{the86} later used this dataset to measure the mass with a modified version of the virial theorem.  \citet{hughes89} used X-ray data under the assumption that the cluster is in hydrostatic equilibrium.  More recently \citet{geller99} used the cluster infall region of Coma to extrapolate the mass out to $10h^{-1}\mathrm{Mpc}$ as well as to determine the virial mass.
   
Weak lensing has become a powerful tool with which to measure the mass of galaxy clusters \citep{tyson90}.  Here the shape distortion of background galaxies due to a foreground cluster is used to determine the mass of the cluster independent of dynamical assumptions.  The shear induced on the background galaxies can be used to probe the mass of the foreground cluster out to large radii.  The majority of galaxy clusters studied with weak lensing lie at redshift $z\geq0.2$ \citep{dahle07}.  However lower redshift clusters have been probed using this method, for instance \citet{joffre00} studied Abell 3667 at $z=0.055$.  At very low redshifts weak lensing measurements of clusters become difficult since this requires imaging of a large area surrounding the cluster.  

The possibility of a weak lensing study of the Coma Cluster $(z=0.0236)$ with the SDSS was first suggested by \citet{gould94} and \citet{stebbins96}.  They pointed out that since Coma is at such a low redshift the weak lensing signal would be measurable in the SDSS since virtually all background galaxies would be at a much higher redshift than Coma.  This is in spite of the shallow imaging in the SDSS which yields a typical background galaxy surface density of $\sim 1$ galaxy $\mathrm{arcmin}^{-2}$ \citep{sheldon04}.  Complete imaging of the entire region surrounding the Coma Cluster was recently completed in the SDSS Data Release Five \citep{adelman07}. 

Here we present a weak lensing measurement of the virial mass of the Coma Cluster in the SDSS.  Our paper is organized as follows: in $\S \ref{sec:analysis}$ we discuss our weak lensing analysis, in $\S \ref{sec:mass}$ our determination of the mass of Coma is discussed, and in $\S \ref{sec:compare}$ we compare our weak lensing mass estimate to previous measurements of the virial mass of Coma. 

Since Coma is at such a low redshift our results do not change significantly for different density parameters.  However for completeness we have assumed a standard cosmology with $\Omega_{m}=0.3$ and $\Omega_{\Lambda}=0.7$.
 
%% In a manner similar to \objectname authors can provide links to dataset
%% hosted at participating data centers via the \dataset{} command.  The
%% second curly bracket argument is printed in the text while the first
%% parentheses argument serves as the valid data set identifier.  Large
%% lists of data set are best provided in a table (see Table 3 for an example).
%% Valid data set identifiers should be obtained from the data center that
%% is currently hosting the data.
%%
%% Note that AASTeX interprets everything between the curly braces in the
%% macro as regular text, so any special characters, e.g. "#" or "_," must be
%% preceded by a backslash. Otherwise, you will get a LaTeX error when you
%% compile your manuscript.  Special characters do not
%% need to be escaped in the optional, square-bracket argument.

%% In this section, we use  the \subsection command to set off
%% a subsection.  \footnote is used to insert a footnote to the text.

%% Observe the use of the LaTeX \label
%% command after the \subsection to give a symbolic KEY to the
%% subsection for cross-referencing in a \ref command.
%% You can use LaTeX's \ref and \label commands to keep track of
%% cross-references to sections, equations, tables, and figures.
%% That way, if you change the order of any elements, LaTeX will
%% automatically renumber them.

%% This section also includes several of the displayed math environments
%% mentioned in the Author Guide.

\section{Weak Lensing Analysis}
\label{sec:analysis}

\subsection{Data}
\label{sec:data}
Data used in our study are obtained from the Sloan Digital Sky Survey (SDSS) \citep{york00} a large imaging and spectroscopic survey of an 8000 square degree region in the Northern Galactic Cap centered on $\alpha=12^{\mathrm{h}}22^{\mathrm{m}}$, $\delta=32^{\circ}13\arcmin$ (J2000).  The SDSS uses a dedicated 2.5m telescope at Apache Point Observatory which images the sky in the $ugriz$ bands \citep{fukugita96} in a drift scan mode \citep{stoughton02}.  The astrometric calibration of the SDSS is described in \citet{pier03} and the photometric calibration is described in \citet{tucker06} and \citet{hogg01}.  For our analysis we use the SDSS Data Release Five \citep{adelman07} which contains complete imaging of the region surrounding Coma. 

Object detection and shape measurement are performed using the PHOTO pipeline \citep{lupton01}.  PHOTO measures a large number of parameters for each detected object in the SDSS \citep{stoughton02}, including adaptive moments \citep{bernstein02} which we use to measure galaxy shapes.  In the SDSS adaptive moments are measured using an iterative algorithm which adapts a Gaussian weight function to the size and shape of each galaxy \citep{sheldon04}.  PHOTO convolves each image with the local PSF before detecting objects in order to avoid the selection bias described in \citet{bernstein02}.  Drift scanning in the SDSS creates a time dependent PSF which leads to a spatial variation in image quality.  To model the PSF in the SDSS a Karhunen-Loeve (KL) decomposition is used, described in further detail in \citet{lupton01} and \citet{sheldon04}.  Camera shear due to drift scanning in the SDSS has been previously shown to be small \citep{hirata04}. 
  
For our weak lensing analysis we use a $\sim200$ square degree region centered on the core of the Coma Cluster.  The core of Coma contains two bright cluster galaxies (BCGs) : NGC 4874 \& NGC 4889.  We take as the center of Coma the galaxy NCG 4874 located at $\alpha=13^{\mathrm{h}}02^{\mathrm{m}}00.2^{\mathrm{s}}$, $\delta=27^{\circ}41^{'}26.6^{''}$ (J2000).  This BCG is associated with the majority of the X-ray flux in the core \citep{vikhlinin01} and the large number of galaxies surrounding this BCG in the optical imaging is indicative of this being the center of the cluster potential (see e.g., \citealt{hansen05}).  We in fact find that the shear amplitude is maximized when we use this BCG as the cluster center.

\subsection{Source Galaxies}
\label{sec:source}
\begin{figure}
\epsscale{1.0}
\plotone{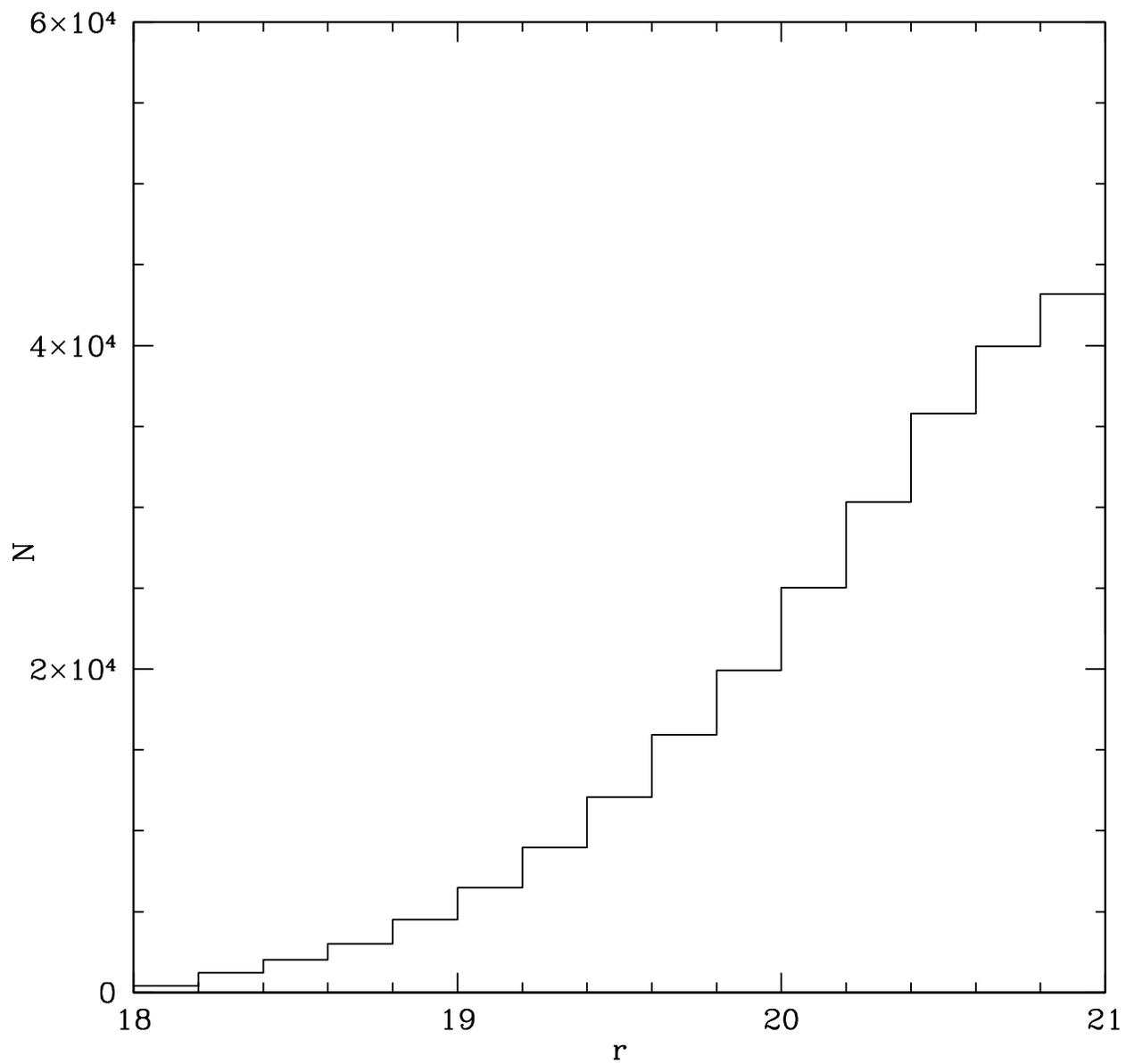}
\caption{r band magnitude distribution of source galaxies used in our analysis. Objects with extinction corrected model magnitudes between $18<r<21$ are used.  The total number of source galaxies used to measure the shear due to Coma is $\sim 270,000$.}
\label{fig:mag}
\end{figure}

To create our source galaxy catalog we use only objects which have been detected in all three $gri$ filters.  We eliminate objects which contain saturated pixels, were originally blended with another object, or triggered errors in the measurement of adaptive moments by rejecting objects with the respective PHOTO error flag set.  For our study we use only objects which have been classified by PHOTO as galaxies $(\mathrm{type}=3)$.  We additionally use only shape measurements from the $r$ band since this filter is the most sensitive and typically contains better seeing \citep{adelman07}.  Galaxies are selected in the $r$ band with extinction corrected model magnitudes \citep{stoughton02} in the range $18<r<21$ (Figure \ref{fig:mag}).  Model magnitudes are used here instead of Petrosian magnitudes since these provide a higher S/N \citep{mandelbaum05}.  Galaxies with fainter magnitudes have been used in galaxy-galaxy weak lensing studies in the SDSS \citep{sheldon04}, but here we restrict our sample to the magnitude range where errors in shape measurements are typically small \citep{hirata04} and galaxy sizes are larger.

In our analysis we use photometric redshifts for our source galaxies.  The SDSS photometric redshift pipeline of \citet{csabai03} is used since it completely covers the area of Coma in Data Release Five.  Galaxies in our magnitude range are typically at much higher redshift than Coma, but we use photometric redshifts here to aid in the rejection of fainter cluster members.  For our study we use only galaxies with photometric redshifts in the range $0.2<z_{\mathrm{phot}}<0.8$ and redshift errors $z_{\mathrm{err}}<0.4$.  Galaxies with lower photometric redshifts are not used as the fractional error in redshift rapidly increases and most of the shear signal comes from galaxies at high redshifts compared to Coma.

To correct for the PSF anisotropy and dilution we use the linear PSF correction scheme of \citet{hirata03}.  This algorithm uses the measured galaxy ellipticity and reconstructed PSF at the position of the object to correct the ellipticity components of each galaxy.  We use galaxies with corrected ellipticities $e_{\mathrm{corr}}<1.4$ in our analysis, which is typical for weak lensing studies in the SDSS \citep{hirata04}.  We additionally make a cut on the resolution parameter $(R)$, defined as  
\begin{equation}R=1-\frac{M^{\mathrm{PSF}}_{\mathrm{rrcc}}}{M_{\mathrm{rrcc}}}\end{equation} \citep{bernstein02} where $M_{\mathrm{rrcc}}$ is the size of the galaxy and $M^{\mathrm{PSF}}_{\mathrm{rrcc}}$ is the size of the PSF at the position of the galaxy as calculated by PHOTO.  For our analysis we use only galaxies with $R>0.33$, which is equivalent to only using objects $1.5$ times larger than the PSF.  We additionally rotate the PSF corrected ellipticity components of each galaxy from image coordinates to the equatorial coordinate system.

Within the annulus described in $\S \ref{sec:shear}$, the total number of galaxies in our source galaxy catalog after applying these cuts is $\sim270,000$.

\subsection{Shear Measurement}
\label{sec:shear}
The shear due to Coma is measured by projecting the PSF corrected ellipticity components of each source galaxy to the tangential frame and binning the galaxies into radial annuli between $0.05\mathrm{h^{-1}Mpc}$ and $10.5\mathrm{h^{-1}Mpc}$.  The edge of the inner annulus is chosen to be larger than the Einstein radius of Coma $(\sim30\arcsec)$ to ensure we are outside of the strong lensing regime, and also to avoid contamination from the BCGs.  Since our radial annuli extend out to $\sim10\mathrm{h^{-1}Mpc}$ ($\simeq8^{\circ}$), we modify the usual flat sky shear matrix into a curved sky shear matrix, following \citet{castro05}.  The tangential shear is measured using
\begin{equation}
\gamma_{t}=\frac{1}{2\mathcal{R}}\frac{\sum e_{t}}{N}
\end{equation}
where $\gamma_{t}$ is the tangential shear, $e_{t}$ is the tangential ellipticity, N is the total number of objects in each radial bin, and $\mathcal{R}$ is the shear responsivity described in \citet{bernstein02}.  We chose not to use any ellipticity error weighting here since our galaxies are selected in a magnitude range where source galaxy ellipticity error is typically small in the SDSS \citep{hirata04}.  We additionally also chose not to weight each source galaxy by the critical surface mass density because all of our source galaxies are at a much higher redshift than Coma and therefore the weights are very nearly equal.  In the case of no weights, the shear responsivity is $\mathcal{R}=1-\sigma_{\mathrm{SN}}^{2}$, where $\sigma_{\mathrm{SN}}$ is the shape noise (the width of the intrinsic ellipticity distribution per ellipticity component) \citep{wittman02}.  Here we use a shape noise of $\sigma_{\mathrm{SN}}=0.37$ which was used previously in \citet{hirata04} for the same source galaxy magnitude range. 

\begin{figure}
\epsscale{1.0}
\plotone{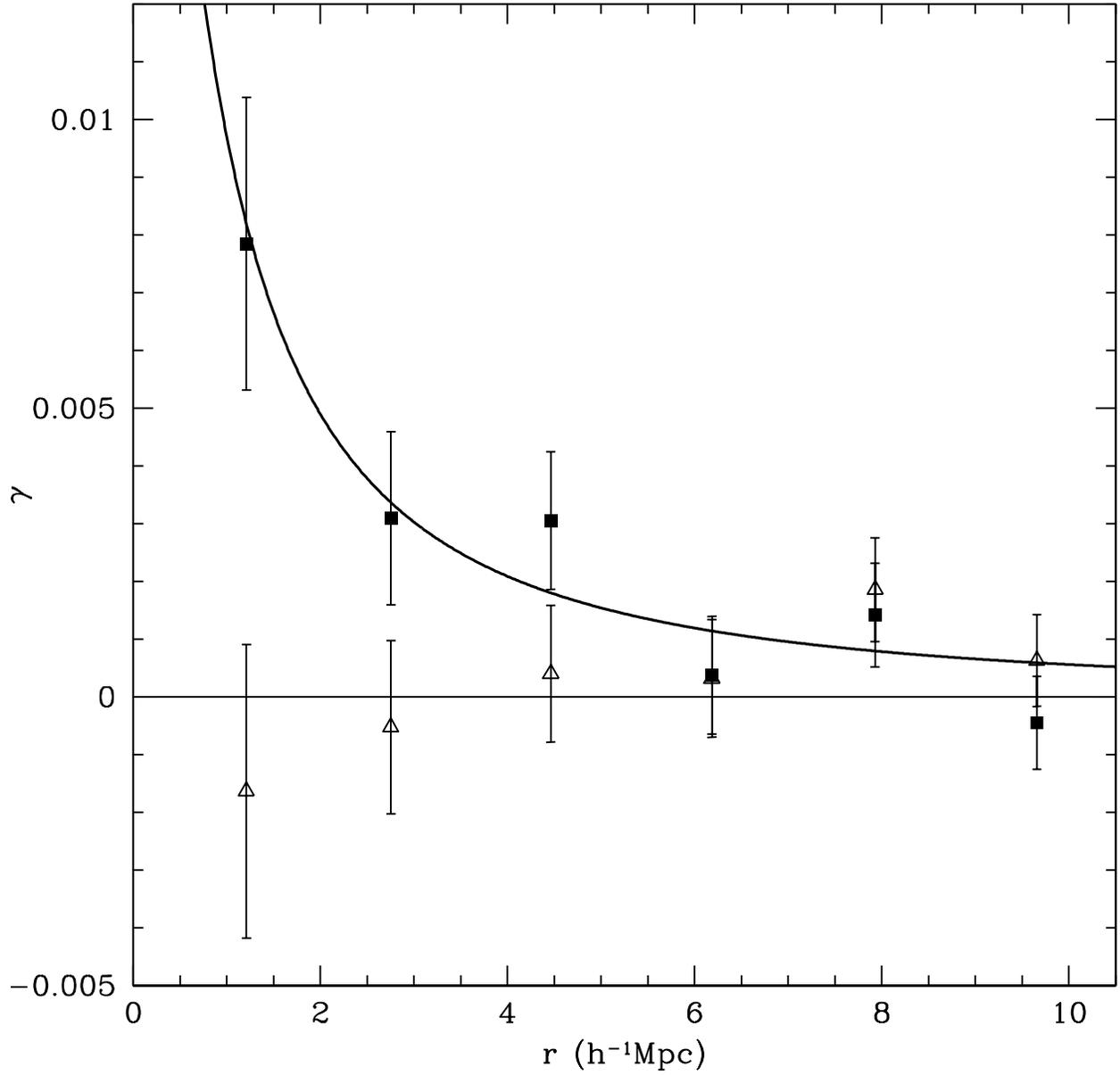}
\caption{Tangential shear centered on the Coma Cluster in the SDSS.  The measured shear is shown as \textit{solid squares} along with $1\sigma$ error bars.  The $45^{\circ}$ component is shown as \textit{open triangles} and is consistent with zero.  The \textit{solid line} represents our best fit NFW model.}
\label{fig:shear}
\end{figure}

The resulting shear profile for Coma is shown in Figure \ref{fig:shear}.  We expect to detect the shear out to $\sim5\mathrm{h^{-1}Mpc}$, and the shear (solid squares) is clearly detected in these inner bins.  Also shown in Figure \ref{fig:shear} is a null test (open triangles) where each source galaxy is rotated by $45^{\circ}$.  The shear signal should dissappear after this rotation if the signal is due to lensing.  Errors shown for both quantities in Figure \ref{fig:shear} are $1\sigma$ error bars, where $\sigma$ is the standard deviation of the mean in each radial bin.  

To statistically test the significance of our measured shear signal we test whether it is consistent with the null shear model ($\gamma_{t}=0$).  We find that our measured shear gives a $\chi^{2}=23.33$ for $6$ degrees of freedom for the null model.  The probability of obtaining a reduced $\tilde{\chi}^{2}$ greater than this value is $\mathrm{P}(\tilde{\chi}^{2}\geq\tilde{\chi}_{o}^{2})=0.07\%$, and therefore we can reject the null hypothesis.  For comparison, fitting the shear from the $45^{\circ}$ test to the null model gives a $\chi^{2}=5.65$ for $6$ degrees of freedom, which is consistent with the null hypothesis.  

\subsection{Blank Fields Test}

\begin{figure}
\epsscale{1.0}
\plotone{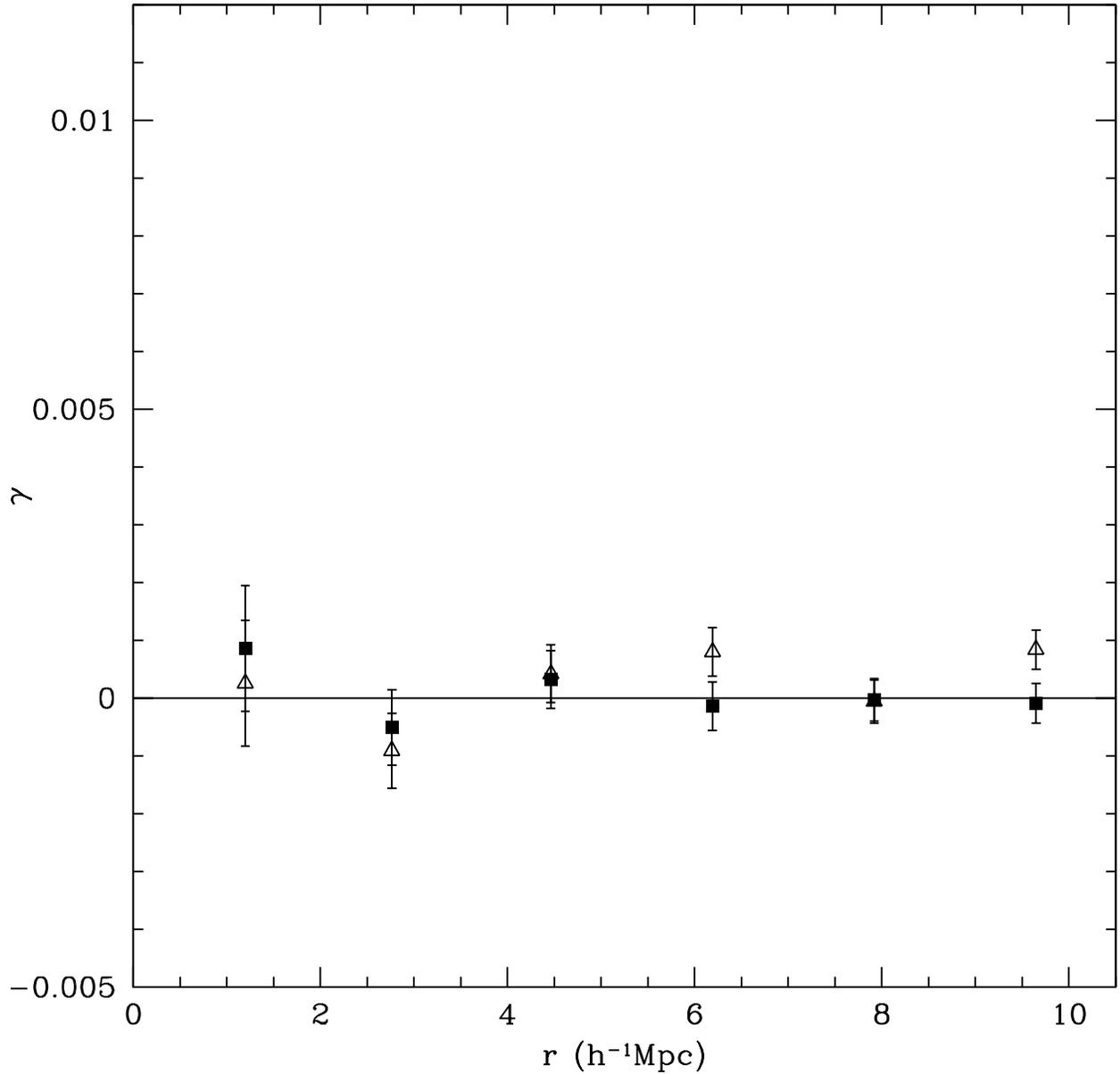}
\caption{Average tangential shear measured from ``blank'' regions of the SDSS North not associated with Coma.  The shear signal (\textit{solid squares}) and $45^{\circ}$ component (\textit{open triangles}) should be zero in these regions.  Results are the inverse variance weighted average of 6 independent $\sim20^{\circ}\times20^{\circ}$ patches.   The size of the annuli here match those used to evaluate the shear due to Coma.}
\label{fig:blank}
\end{figure}

To test that our signal is not affected by any remaining systematic in the survey, we split up the SDSS North into separate, non-overlapping `Coma sized' patches that contained no galaxies used in our Coma analysis and which contained no large sections of missing data.  We were able to successfully extract 6 blank fields in DR5 each $\sim 20^{\circ}\times20^{\circ}$ wide.  From the center of each patch we probed radially outward to $\sim10\mathrm{h^{-1}Mpc}$ at the redshift of Coma.  We applied the same cuts to the source galaxy catalog in each of the blank fields that were used in our Coma analysis in $\S \ref{sec:source}$.  The resulting inverse variance weighted average signal over all fields is shown in Figure \ref{fig:blank}, where we have used the same binning as in our Coma analysis.  The tangential shear signal is shown with solid squares and the $45^{\circ}$ component is shown with open triangles.  The tangential shear component is consistent with the null model giving a $\chi^{2}=1.84$ for $6$ degrees of freedom.  The $45^{\circ}$ component gives a $\chi^{2}=12.58$ for $6$ degrees of freedom which has a probability $\mathrm{P}(\tilde{\chi}^{2}\geq\tilde{\chi}_{o}^{2})=5.0\%$.  A model with a small positive shear is slightly favored for the $45^{\circ}$ component, however the null model can only be excluded at the $5\%$ level.  The field to field scatter in the blank field shear measurements shows a typical standard deviation in a given radial bin of $\sigma_{\gamma}\sim0.0015$ or a standard deviation of the mean of $\sim 0.0006$, similar in magnitude to the statistical errors that are plotted in Figure \ref{fig:blank}.  In principle the scatter over the blank fields can provide information on the error due to large scale structure (\citealt{hoekstra01}; also \S \ref{sec:error} below), but our determination of the scatter is limited by the finite number of blanks fields.

%% Putting eqnarrays or equations inside the mathletters environment groups
%% the enclosed equations by letter. For instance, the eqnarray below, instead
%% of being numbered, say, (4) and (5), would be numbered (4a) and (4b).
%% LaTeX the paper and look at the output to see the results.

%% This section contains more display math examples, including unnumbered
%% equations (displaymath environment). The last paragraph includes some
%% examples of in-line math featuring a couple of the AASTeX symbol macros.

\section{Mass Model}
\label{sec:mass}

\subsection{Tangential Shear}

The tangential shear due to a foreground cluster lens is given by 
\begin{equation}\gamma_{t}=\frac{\bar{\Sigma}(\leq r)-\Sigma(r)}{\Sigma_{\mathrm{crit}}}\end{equation}
where $\bar{\Sigma}(\leq r)$ is the average projected surface mass density interior to $r$, and $\Sigma(r)$ is the projected surface mass density at $r$ \citep{miralda91}.  The magnitude of the shear also depends on the critical surface mass density $\Sigma_{\mathrm{crit}}$ which is determined by 
\begin{equation}\Sigma_{\mathrm{crit}}=\frac{c^2}{4\pi G}\frac{D_{s}}{D_{l}D_{ls}}.\end{equation}
Here $D_{l}$ and $D_{s}$ are the angular diameter distances from the observer to the lens and source respectively, and $D_{ls}$ is the angular diameter distance between the lens and source.  

To compute the critical surface mass density we use for the lens the exact spectroscopic redshift of Coma $z=0.0236$ \citep{geller99}.  We assume here that the peculiar velocity for Coma is zero, which has been shown in several studies \citep{scodeggio97,giovanelli97}.  Photometric redshifts are used to calculate the source angular diameter distances, and we obtain a critical surface mass density of $26299 \mathrm{M_{\odot} pc^{-2}}$.  For comparison this is nearly the value obtained if we assume all of our sources were at a fixed redshift of $z=0.3$.

\subsection{NFW Profile}
\label{sec:nfw}

Since the S/N of our data is low $(S/N\sim 5)$ we chose to fit the measured shear profile of Coma to a model.  We fit the shear to that expected from a Navarro, Frenk, \& White (NFW) profile \citep{navarro96}.  The NFW profile is a ``universal profile'' found in N-body simulations to fit mass density profiles ranging from galaxies to galaxy clusters.  The density of an NFW profile is described by
\begin{equation}
\rho(r)=\frac{\delta_{c}\rho_{c}}{(r/r_{s})(1+r/r_{s})^2}
\end{equation}
where $\delta_{c}$ is the halo overdensity, $r_{s}$ is the scale radius, and $\rho_{c}=3H^{2}(z)/8\pi G$ is the critical density at the redshift of the cluster.  The halo overdensity is given by
\begin{equation}\delta_{c}=\frac{200}{3}\frac{c^3}{\mathrm{ln}(1+c)-c/(1+c)}\end{equation}
where $c$ is the halo concentration.  The NFW model is therefore determined by only two parameters $c$ and $r_{s}$ which are highly correlated.  The NFW model can be used to determine the virial radius $r_{200}=cr_{s}$ defined as the radius in which the interior mass density falls to $200\rho_{c}$.  The corresponding virial mass $M_{200}$ is given by 
\begin{equation}M_{200}=\frac{800\pi}{3}\rho_{c}r_{200}^{3}.\end{equation}

The expected shear due to an NFW profile has been worked out in detail by \citet{wright00} and we use their result in our analysis.  Because the NFW profile is highly non-linear, we use the Levenberg-Marquardt fitting procedure in \citet{press95} to determine the virial radius and the halo concentration.  We find that the values which minimize $\chi^{2}$ are given by
\begin{equation}r_{200}=1.99_{-0.22}^{+0.21}\mathrm{h}^{-1}\mathrm{Mpc}\end{equation}
and
\begin{equation}c=3.84_{-1.84}^{+13.16}.\end{equation}
The concentration parameter is not well constrained.  This fit gives a $\chi^{2}=3.87$ for 4 degrees of freedom which has a $\mathrm{P}(\tilde{\chi}^{2}\geq\tilde{\chi}_{o})=42.4\%$ probability of occuring.  This corresponds to a virial mass of 
\begin{equation}M_{200}= 1.88_{-0.56}^{+0.65}\times10^{15}\mathrm{M_{\odot}}.\end{equation}
Our values of $r_{200}$ and $M_{200}$ remain constant if the number of bins is slightly changed, and therefore our choice of binning does not bias this result.  It has been shown previously that large scale structure affects the errors in the fit parameters of weak lensing measurements \citep{hoekstra03}, however this is not considered in our fit errors.  We discuss additional sources of systematic error not included in our fit in $\S \ref{sec:error}$.

%% The displaymath environment will produce the same sort of equation as
%% the equation environment, except that the equation will not be numbered
%% by LaTeX.

%% If you wish to include an acknowledgments section in your paper,
%% separate it off from the body of the text using the \acknowledgments
%% command.

%% Included in this acknowledgments section are examples of the
%% AASTeX hypertext markup commands. Use \url without the optional [HREF]
%% argument when you want to print the url directly in the text. Otherwise,
%% use either \url or \anchor, with the HREF as the first argument and the
%% text to be printed in the second.

\section{Discussion}
\label{sec:compare}
We compare our weak lensing mass estimate with three other methods used to determine the virial mass of Coma: (1) The Virial Theorem (2) X-ray  and (3) the Infall Region.  Our value for $M_{200}$ is consistently higher than the previous measurements, though never by more that $2\sigma$.  These comparisons are summarized in Table \ref{tab:mass}.  We also discuss sources of additional error not included in our mass measurement, as well as the source of any potential bias.

\subsection{Virial Theorem}
\citet{the86} used a sample of galaxies in Coma along with the virial theorem to measure its mass.  Their analysis also accounted for an additional portion of the system that may have been missing in their spectroscopic sample.  Relaxing the assumption that light traces mass in the system, they found that the mass varied over a wide range.  However, assuming that the mass traces the galaxy distribution they found a mass within $2.7\mathrm{h^{-1}Mpc}$ of $0.95\times10^{15}\mathrm{h^{-1}M_{\odot}}$ with a $15\%$ error in mass.  Our weak lensing measurement is consistent with this result to within $\sim1.6\sigma$.

\subsection{X-ray}
Our weak lensing measurement is also consistent with the previous X-ray analysis of \citet{hughes89}.  This study used the assumption of hydrostatic equilibrium but avoided certain assumptions about the state of the gas and also accounted for many systematic effects.  The best fit model was one in which mass traces light, giving a virial mass of $(0.93\pm0.12)\times 10^{15} \mathrm{h^{-1} M_{\odot}}$ within $2.5\mathrm{h^{-1}}\mathrm{Mpc}$ which is within $\sim1.7\sigma$ of our result. 
 
\subsection{Infall Region}
Using a redshift survey of $\sim 1000$ galaxies \citet{geller99} used the infall region of Coma to determine its mass.  In redshift space galaxies falling into the potential well of the cluster are used to determine the amplitude of the redshift caustics (the boundaries in line of sight velocity vs. projected radius) \citep{diaferio97}.  The amplitude of the caustics along with the assumptions of spherical symmetry and hierarchal clustering are directly related to the cluster gravitational potential.  Using this technique they find an NFW model fits the mass well giving an estimate for $r_{200}=1.5\mathrm{h^{-1}}\mathrm{Mpc}$ which yields an $M_{200}=0.8\times10^{15}M_{\odot}$.  Error bars are not quoted for their virial radius or mass; however, our weak lensing virial mass is in agreement with their central result to within $\sim2.0\sigma$

\subsection{Additional Error and Bias}
\label{sec:error}
Weak lensing measurements of the mass of clusters, in particular low redshift clusters, suffer from two sources of additional error: (1) distant large scale structure and (2) correlated structure near the cluster \citep{hoekstra03}.  Weak lensing is sensitive to all structure along the sight and it has been shown that any background structure introduces additional error in the weak lensing mass estimate of clusters \citep{hoekstra01}.  Specifically for the case of the Coma Cluster, \citet{hoekstra01} showed that the error due to large scale structure increases with imaging depth, thus limiting the total achievable S/N for Coma to $\sim7$.  In our study the imaging depth in SDSS is relatively shallow so the contribution of large scale structure to the total error budget is small.  For future deep imaging surveys which will also cover the area of Coma such as PanSTARRS \citep{kaiser02}, the error due to large scale structure will be more significant. 

The effect of correlated structure (such as filaments) near a foreground cluster has been previously studied using N-body simulations by \citet{cen97} and \citet{metzler99}.  Here it was shown that filaments can increase the statistical error in weak lensing mass estimates as well as cause the measured $M_{200}$ to be biased toward higher values.  In principle Coma should be an interesting testing ground to study this effect since the neighboring structure is well understood.  However in practice there is not a clear method to correct a weak lensing measured value of $M_{200}$ for this effect, and therefore we leave a study of any potential bias in our measurement for a future paper. 

\section{Conclusion}
\label{sec:end}
We have measured the weak lensing shear due to the Coma Cluster, currently the lowest redshift $(z=0.0236)$ largest angle weak lensing measurement of an individual cluster to date.  Our analysis is performed using the SDSS which is the only imaging survey that covers a large enough area to measure the shear due to Coma.  We find the shear can be fit well to an NFW profile and have compared our weak lensing derived mass estimate to dynamical methods which have been used to probe the mass of Coma out to large radius.  In particular we have compared our weak lensing mass to the mass derived using the virial theorem \citep{the86}, X-ray data \citep{hughes89}, and the cluster infall region \citep{geller99}. 
We find the virial mass of Coma from weak lensing is consistent with the mass derived using these other techniques, though the error from weak lensing is larger.

\acknowledgments

We thank the Fermilab Clusters Group for useful comments during the course of this work.  Funding for the SDSS and SDSS-II has been provided by the Alfred P. Sloan Foundation, the Participating Institutions, the National Science Foundation, the U.S. Department of Energy, the National Aeronautics and Space Administration, the Japanese Monbukagakusho, the Max Planck Society, and the Higher Education Funding Council for England. The SDSS Web Site is http://www.sdss.org/.

    The SDSS is managed by the Astrophysical Research Consortium for the Participating Institutions. The Participating Institutions are the American Museum of Natural History, Astrophysical Institute Potsdam, University of Basel, University of Cambridge, Case Western Reserve University, University of Chicago, Drexel University, Fermilab, the Institute for Advanced Study, the Japan Participation Group, Johns Hopkins University, the Joint Institute for Nuclear Astrophysics, the Kavli Institute for Particle Astrophysics and Cosmology, the Korean Scientist Group, the Chinese Academy of Sciences (LAMOST), Los Alamos National Laboratory, the Max-Planck-Institute for Astronomy (MPIA), the Max-Planck-Institute for Astrophysics (MPA), New Mexico State University, Ohio State University, University of Pittsburgh, University of Portsmouth, Princeton University, the United States Naval Observatory, and the University of Washington.

\clearpage

%% Use the figure environment and \plotone or \plottwo to include
%% figures and captions in your electronic submission.
%% To embed the sample graphics in
%% the file, uncomment the \plotone, \plottwo, and
%% \includegraphics commands
%%
%% If you need a layout that cannot be achieved with \plotone or
%% \plottwo, you can invoke the graphicx package directly with the
%% \includegraphics command or use \plotfiddle. For more information,
%% please see the tutorial on "Using Electronic Art with AASTeX" in the
%% documentation section at the AASTeX Web site,
%% http://www.journals.uchicago.edu/AAS/AASTeX.
%%
%% The examples below also include sample markup for submission of
%% supplemental electronic materials. As always, be sure to check
%% the instructions to authors for the journal you are submitting to
%% for specific submissions guidelines as they vary from
%% journal to journal.

%% This example uses \plotone to include an EPS file scaled to
%% 80% of its natural size with \epsscale. Its caption
%% has been written to indicate that additional figure parts will be
%% available in the electronic journal.

\begin{deluxetable}{ccc} 
\tablecolumns{3} 
\tablewidth{0pt}
\tablecaption{Virial Mass Estimates of Coma} 
\tablehead{ 
\colhead{Virial Radius}  & \colhead{Mass} &\colhead{Reference}\\
\colhead{($\mathrm{h^{-1}Mpc}$)} & \colhead{($\mathrm{10^{15}h^{-1}M_{\odot}}$)}
}
\startdata 
$1.99_{-0.22}^{+0.21}$   & $1.88_{-0.56}^{+0.65}$ & this work\tablenotemark{a}\\ 
$1.5$ & $0.8$ & 1\tablenotemark{a}\\
$2.5$ & $0.93\pm0.12$ &2\\
$2.7$ & $0.95\pm0.15$ &3\\
\enddata 
\tablenotetext{a}{Determined from the NFW $r_{200}$ virial radius.}
\tablerefs{(1) \citealt{geller99}; (2) \citealt{hughes89}; (3) \citealt{the86}.}
\label{tab:mass}
\end{deluxetable} 

\clearpage

%% Here we use \plottwo to present two versions of the same figure,
%% one in black and white for print the other in RGB color
%% for online presentation. Note that the caption indicates
%% that a color version of the figure will be available online.
%%

%% If you are not including electonic art with your submission, you may
%% mark up your captions using the \figcaption command. See the
%% User Guide for details.
%%
%% No more than seven \figcaption commands are allowed per page,
%% so if you have more than seven captions, insert a \clearpage
%% after every seventh one.

%% Tables should be submitted one per page, so put a \clearpage before
%% each one.

%% Two options are available to the author for producing tables:  the
%% deluxetable environment provided by the AASTeX package or the LaTeX
%% table environment.  Use of deluxetable is preferred.
%%

%% Three table samples follow, two marked up in the deluxetable environment,
%% one marked up as a LaTeX table.

%% In this first example, note that the \tabletypesize{}
%% command has been used to reduce the font size of the table.
%% We also use the \rotate command to rotate the table to
%% landscape orientation since it is very wide even at the
%% reduced font size.
%%
%% Note also that the \label command needs to be placed
%% inside the \tablecaption.

%% This table also includes a table comment indicating that the full
%% version will be available in machine-readable format in the electronic
%% edition.

\clearpage

%% Text for table notes should follow after the \enddata but before
%% the \end{deluxetable}. Make sure there is at least one \tablenotemark
%% in the table for each \tablenotetext.

%% If you use the table environment, please indicate horizontal rules using
%% \tableline, not \hline.
%% Do not put multiple tabular environments within a single table.
%% The optional \label should appear inside the \caption command.

\clearpage

%% If the table is more than one page long, the width of the table can vary
%% from page to page when the default \tablewidth is used, as below.  The
%% individual table widths for each page will be written to the log file; a
%% maximum tablewidth for the table can be computed from these values.
%% The \tablewidth argument can then be reset and the file reprocessed, so
%% that the table is of uniform width throughout. Try getting the widths
%% from the log file and changing the \tablewidth parameter to see how
%% adjusting this value affects table formatting.

%% The \dataset{} macro has also been applied to a few of the objects to
%% show how many observations can be tagged in a table.

\clearpage

%% The following command ends your manuscript. LaTeX will ignore any text
%% that appears after it.


\begin{thebibliography}{99}
\bibitem[Adelman-McCarthy et al.(2007)]{adelman07}Adelman-McCarthy, J., et al. 2007, arXiv:0707.3380
\bibitem[Bernstein \& Jarvis(2002)]{bernstein02}Bernstein, G.M. \& Jarvis, M.J. 2002, \aj, 123, 583
\bibitem[Castro et al.(2005)]{castro05}Castro, P.G., Heavens, A.F., Kitching, T.D. 2005, Phys. Rev. D, 72, 023516
\bibitem[Cen(1997)]{cen97}Cen, R. 1997, \apj, 485, 39
\bibitem[Dahle(2007)]{dahle07}Dahle, H. 2007, preprint (astro-ph/0701598)
\bibitem[Diaferio \& Geller(1997)]{diaferio97}Diaferio, A., \& Geller, M.J. 1997, \apj, 481, 633
\bibitem[Csabai et al.(2003)]{csabai03}Csabai, I., et al. 2003, \aj, 125, 580
\bibitem[Fukugita et al.(1996)]{fukugita96}Fukugita, M., Ichikawa, T., Gunn, J.E., Doi, M., Shimasaku, K., \& Schneider, D.P. 1996, \aj, 111, 1748
\bibitem[Geller et al.(1999)]{geller99} Geller, M. J., Diaferio, A., Kurtz, M.J. 1999, \apj, 517L, 23
\bibitem[Giovanelli et al.(1997)]{giovanelli97}Giovanelli, R., Haynes, M.P., Herter, T., da Costa, L.N., Wolfram, F., Salzer, J.J., Wegner, G., 1997, \aj, 113, 53
\bibitem[Gould \& Villumsen(1994)]{gould94}Gould, A. \& Villumsen, J. 1994, \apj, 428L, 45
\bibitem[Hansen et al.(2005)]{hansen05}Hansen, S.M., McKay, T.A., Wechsler, R.H., Annis, J., Sheldon, E.S., Kimball, A. 2005, \apj, 633, 122
\bibitem[Hirata et al.(2004)]{hirata04}Hirata, C.M. 2004, \mnras, 353, 529
\bibitem[Hirata \& Seljak(2003)]{hirata03}Hirata, C.M. \& Seljak, U. 2003, \mnras, 343, 459
\bibitem[Hoekstra(2001)]{hoekstra01}Hoekstra, H. 2001, A\&A, 370, 743
\bibitem[Hoekstra(2003)]{hoekstra03}Hoekstra, H. 2003, \mnras, 339, 1155
\bibitem[Hogg et al.(2001)]{hogg01}Hogg, D.W., Finkbeiner, D.P., Schlegel, D.J., \& Gunn, J.E. 2001, \aj, 122, 2129
\bibitem[Hughes(1989)]{hughes89}Hughes, J.P, 1989, \apj, 337, 21
\bibitem[Joffre et al.(2000)]{joffre00}Joffre, M. et al. 2000, \apj, 534L, 131
\bibitem[Kaiser et al.(2002)]{kaiser02}Kaiser, N. et al. 2002, SPIE, 4836, 154
\bibitem[Kent \& Gunn(1982)]{kent82}Kent, S.M. \& Gunn, J.E. 1982, \aj, 87, 945
\bibitem[Lupton et al.(2001)]{lupton01}Lupton, R., Gunn, J., Ivezic, Z., Knapp, G.R. 2001, preprint (astro-ph/0101420)
\bibitem[Mandelbaum et al.(2005)]{mandelbaum05}Mandelbaum, R., et al. 2005, \mnras, 361, 1287
\bibitem[Metzler et al.(1999)]{metzler99}Metzler, C.A., White, M., Norman, M., Loken, C. 1999, \apj, 520L, 9
\bibitem[Miralda-Escude(1991)]{miralda91}Miralda-Escude, J. 1991, \apj, 370, 1
\bibitem[Navarro et al.(1996)]{navarro96}Navarro, J.F., Frenk, C.S., White, S.D. 1996, \apj, 462, 563
\bibitem[Pier et al.(2003)]{pier03}Pier, J.R., et al. 2003, \aj, 125, 1559
\bibitem[Press et al.(1995)]{press95}Press, W.H., Flannery, B.P., Teukolsky, S.A., Vetterling, W.T. 1995, Numerical Recipes in C (2nd ed.; Cambridge:Cambridge Univ. Press) 
\bibitem[Scodeggio et al.(1997)]{scodeggio97}Scodeggio, M., Giovanelli, R., \& Haynes, M.P., 1997, \aj, 113, 101
\bibitem[Sheldon et al.(2004)]{sheldon04}Sheldon, E.S., 2004, \aj, 127, 2544
\bibitem[Stebbins et al.(1996)]{stebbins96}Stebbins, A., Mckay, T., Frieman, J.A., 1996, in IAU Symp. 173, Astrophysical Applications of Gravitational Lensing, ed.s C.S. Kochanek and J.N. Hewitt (Kluwer Academic, New York), 75
\bibitem[Stoughton et al.(2002)]{stoughton02}Stoughton, C., et al. 2002, \aj, 123, 485
\bibitem[The \& White(1986)]{the86}The, L.S., \& White, S.D. 1986, \aj, 92, 1248
\bibitem[Tucker et al.(2006)]{tucker06}Tucker, D.L., et al., 2006, AN, 327, 821
\bibitem[Tyson et al.(1990)]{tyson90}Tyson, J.A., Wenk, R.A., \& Valdes, F. 1990, \apj, 349L, 1
\bibitem[Vikhlinin et al.(2001)]{vikhlinin01}Vikhlinin, A., Markevitch, M., Forman, W., Jones, C. 2001, \apj, 555L, 87
\bibitem[Wittman(2002)]{wittman02}Wittman,D. 2002, LNP, 608, 55
\bibitem[Wright \& Brainerd(2000)]{wright00}Wright, C.O. \& Brainerd, T.G. 2000, \apj, 534, 34
\bibitem[York et al.(2000)]{york00}York, D.G., et al. 2000, \aj, 120, 1579
\bibitem[Zwicky(1933)]{zwicky33}Zwicky, F. 1933, Helvetica Phys. Acta, 6, 110
\end{thebibliography}
\end{document}